\documentclass[12pt,english]{paper}
\usepackage{amsthm,amsmath,amssymb}
\usepackage[T1]{fontenc}
\usepackage[latin1]{inputenc}
\usepackage{graphicx}
\usepackage{cite}
\usepackage{slashed}
\makeatletter

\newcommand{\bibstyle@aas}{\bibpunct{(}{)}{;}{a}{}{,}}

\makeatletter

\makeatletter

\usepackage{geometry}

\makeatletter

\usepackage{epsfig}

\makeatother

\makeatother

\makeatother

\usepackage{babel}
\makeatother
\begin{document}

\title{Search for Milli-Charged Particles from the Sun at IceCube}

\author{Ye Xu$^{1,2}$}

\maketitle

\begin{flushleft}
$^1$School of Electronic, Electrical Engineering and Physics,  Fujian University of Technology, Fuzhou 350118, China
\par
$^2$Research center for Microelectronics Technology, Fujian University of Technology, Fuzhou 350118, China
\par
e-mail address: xuy@fjut.edu.cn
\end{flushleft}

\begin{abstract}
It is assumed that heavy dark matter particles $\phi$ with O(TeV) mass captured by the Sun may decay to relativistic light milli-charged particles (MCPs). These MCPs could be measured by the IceCube detector. The massless hidden photon model was taken for MCPs to interact with nuclei, so that the numbers and fluxes of expected MCPs and neutrinos may be evaluated at IceCube. Based on the assumption that no events are observed at IceCube in 6 years, the corresponding upper limits on MCP fluxes were calculated at 90\% C. L.. These results indicated that MCPs could be directly detected in the secondaries' energy range O(100GeV)-O(10TeV) at IceCube, when $\epsilon^2\gtrsim10^{-10}$. And a new region of 0.6 MeV < $m_{MCP}$ < 10 MeV and $6\times10^{-6}$ < $\epsilon$ $\lesssim$ $10^{-4}$ is ruled out in the $m_{MCP}$-$\epsilon$ plane with 6 years of IceCube data.
\end{abstract}

\section{Introduction}
It was found in cosmological and astrophysical observations that most ($84\%$) of matter in the Universe consists of dark matter (DM)\cite{bergstrom,BHS,Planck2015}. So far, DM has been observed only through its gravitational interactions. DM is neutral under all Standard Model (SM) gauge interactions in most of DM models, for example, weak interacting massive particles (WIMPs), axions, axion-like particles, sterile neutrinos and so on. Unfortunately, no one has found those neutral DM particles yet\cite{XENON1T,PANDAX,fermi,antares-icecube-dm-MW,icecubedm-sun,antaresdm-sun,CAST,GlueX,NGC1275,Chooz,dayabayMINOS}.
\par
Milli-charged particles (MCPs), which are fermions, with a small electric charge $\epsilon e$ ($e$ is the electric charge for an electron  and $\epsilon\ll1$), are an alternative DM scenario\cite{GH,CY,FLN}. A model with a hidden gauge group U(1) is taken for MCPs to interact with nuclei. A second unbroken "mirror" U(1)$^{\prime}$ was introduced in this model. The corresponding massless hidden photon field may have a kinetic mixing to the SM photon, so that a MCP under U(1)$^{\prime}$ appears to have a small coupling to the SM photon\cite{Holdom}. Certainly, MCPs can also arise in extra-dimensional scenarios or as hidden magnetic monopoles receiving their mass from a magnetic mixing effect\cite{BG,BJ,BJK}. $\epsilon$ is also the kinetic mixing parameter between those two kinds of photons. The searches for MCPs have been performed in cosmological and astrophysical observations, accelerator experiments, experiments for decay of ortho-positronium and Lamb shift, DM searches and so on, so that constraints on $\epsilon$ were determined by those observations. \cite{CM,DHR,GH,DCB,SLAC,Xenon,LS,OP,Lamb}.
\par
In the DM scenario in this work, there exist at least two DM species in the Universe (for example, O(TeV) DM and light MCPs). O(TeV) DM, $\phi$, is a thermal particle which is generated by the early universe. The bulk of present-day DM consists of them. The other is a stable light fermion, MCP ($\chi$), which is the product of the decay of $\phi$ ($\phi\to\chi\bar{\chi}$), like the DM decay channel mentioned in Ref.\cite{FR}. It is assumed that its mass is much less than that of a proton. Due to the decay of long-living $\phi$ ($\tau_{\phi} \gg t_0$\cite{AMO,EIP}, $t_0\sim10^{17}$ s is the age of the Universe. Here $\tau_{\phi} \geq 10^{19}$ s), the present-day DM may also contain a very small component which is MCPs with the energy of about $\displaystyle\frac{m_\phi}{2}$. Here it is assumed that the decay of $\phi$ are only through $\phi\to\chi\bar{\chi}$.
\par
The $\phi$'s of the Galactic halo would be captured by the Sun when their wind sweeps through the Sun. The measurement of light neural DM due to the decay of heavy $\phi$ captured by the Sun at IceCube has been discussed in my previous work\cite{xu}. The $Z^{\prime}$-portal model was taken for those neural particles to interact with nuclei. In this work, however, the $\phi$'s captured by the Sun can only decay into MCPs. A model with a massless hidden photon will be taken for MCPs to interact with nuclei. MCPs would interact with nuclei when they pass through the Sun, the Earth and ice. 
Those MCPs can be directly measured with the IceCube neutrino telescope via the deep inelastic scattering (DIS) with nuclei in the ice. The capability of the measurement of those particles will also be discussed here. In this measurement, the background consists of muons and neutrinos generated in cosmic ray interactions in the Earth's atmosphere and astrophysical neutrinos.
\section{Flux of MCPs which reach the Earth}
The $\phi$'s of the Galactic halo would collide with atomic nuclei in the Sun and be captured when their wind sweeps through the Sun. Those $\phi$'s inside the Sun can decay into MCPs at an appreciable rate. Then the number of those $\phi$'s is obtained in the way in Ref.\cite{BCH}
\begin{center}
\begin{equation}
\frac{dN}{dt}=C_{cap}-2\Gamma_{ann}-C_{evp}N-C_{dec}N
\end{equation}
\end{center}
where $C_{cap}$, $\Gamma_{ann}$ and $C_{evp}$ are the capture rate, the annihilation rate and the evaporation rate, respectively. The evaporation rate is only relevant when the DM mass < 5 GeV\cite{BCH}, which are much lower than my interested mass scale (the mass of $\phi$, m$_{\phi}$ $\geq$ 1 TeV). Thus their evaporation contributes to the accumulation in the Sun at a negligible level in the present work. $C_{dec}$ is the decay rate for $\phi$'s. Since the fraction of $\phi$ decay $\leq$ 3.0$\times 10^{-12}$ per year ($\tau_{\phi} \geq 10^{19}$ s), its contribution to the $\phi$ accumulation in the Sun can be ignored in the evaluation of $\phi$ accumulation. $\Gamma_{ann}$ is obtained by the following equation\cite{BCH}
\par
\begin{center}
\begin{equation}
\Gamma_{ann}=\frac{C_{cap}}{2}tanh^2\left(\frac{t}{\tau}\right)\approx \frac{C_{cap}}{2} \quad with \quad t\gg\tau
\end{equation}
\end{center}
where $\tau=(C_{cap}C_{ann})^{-\frac{1}{2}}$ is a time-scale set by the competing processes of capture and annihilation. At late times $t\gg\tau$ one can approximate tanh$^2\displaystyle\frac{t}{\tau}$=1 in the case of the Sun\cite{BCH}. $C_{cap}$ is proportional to $\displaystyle\frac{\sigma_{\phi N}}{m_{\phi}}$\cite{BCH,JKK}, where $m_{\phi}$ is the mass of $\phi$ and $\sigma_{\phi N}$ is the scattering cross section between the nuclei and $\phi$'s. The spin-independent cross section is only considered in the capture rate calculation. Then $\sigma_{\phi N}$ is taken to be 10$^{-44}$ cm$^2$ for $m_{\phi} \sim$ O(TeV) \cite{XENON1T,PANDAX}. Besides, one knows that $\phi$'s are concentrated around the center of the Sun from Ref.\cite{BCH}.
\par
The MCPs which reach the Earth are produced by the decay of $\phi$'s in the Sun's core. Those MCPs have to pass through the Sun and interact with nuclei inside the Sun. Then the number N$_s$ of MCPs which reach the Sun's surface is obtained by the following equation:
\begin{center}
\begin{equation}
\begin{aligned}
N_s &=2N_0\left(exp(-\frac{t_0}{\tau_{\phi}})-exp(-\frac{t_0+T}{\tau_{\phi}})\right)\prod_{i=1}^{n=\mathcal{N}} exp(-\frac{\delta L}{L_i}) \qquad with \quad T \ll \tau_{\phi}\\
    &\approx 2N_0\frac{T}{\tau_{\phi}}exp(-\frac{t_0}{\tau_{\phi}})\prod_{i=1}^{n=\mathcal{N}} exp(-\frac{\delta L}{L_i})
\end{aligned}
\end{equation}
\end{center}
where N$_0$=$\displaystyle\int^{t_s}_0 \displaystyle\frac{dN}{dt} dt$ is the number of $\phi$'s captured in the Sun. t$_s$ and t$_0$ are the ages of the Sun and the Universe, respectively. T is the lifetime of taking data for IceCube and taken to be 6 years. If the distance from the Sun's center to the Sun's surface is equally divided into $\mathcal{N}$ portions, $\delta L=\displaystyle\frac{R_{sun}}{\mathcal{N}}$. $L_i=\displaystyle\frac{1}{N_A\rho_i\sigma_{\chi N}}$ is the MCP interaction length at i$\times\delta L$ away from the Sun's center. $\rho_i$ is the density at i$\times\delta L$ away from the Sun's center\cite{SN}. N$_s$ is computed in column density in the present work. The first exponential term  
in Eqn. (3) is the fraction of decay of $\phi$'s in the Sun's core. The term of continued product 
in Eqn. (3) is the faction of MCPs which reach the Sun's surface. Here $\mathcal{N}$ is taken to be 10$^4$. The results with $\mathcal{N}$=10$^4$ is sufficiently accurate, whose uncertainty is about 0.05\%.
\par
Then the flux $\Phi_{MCP}$ of MCPs, which reach the Earth, from the Sun's core is described by
\begin{center}
\begin{equation}
\Phi_{MCP}=\frac{N_s}{4\pi D_{se}^2}
\end{equation}
\end{center}
where $D_{se}$ is the distance between the Sun and Earth.
\section{MCP and neutrino interactions with nuclei}
In this work, the hidden photon model\cite{Holdom} is taken for MCPs to interact with nuclei via a neural current (NC) interaction mediated by the mediator generated by the kinetic mixing between the SM and massless hidden photons. There is only a well-motivated interaction allowed by SM symmetries that provide a "portal" from the SM particles into the MCPs. This portal is $\displaystyle\frac{\epsilon}{2}F_{\mu\nu}F^{\prime\mu\nu}$. Then its interaction Lagrangian can be written as follows:
\begin{center}
\begin{equation}
\mathcal{L} =\sum_qe_q\bar{q}\gamma^{\mu}qA_{\mu} -\frac{1}{4}F^{\prime}_{\mu\nu}F^{\prime\mu\nu}+\bar{\chi}(i\slashed{D}-m_{\chi})\chi-\frac{\epsilon}{2}F_{\mu\nu}F^{\prime\mu\nu}
\end{equation}
\end{center}
where the sum runs over quark flavors in the nucleon and $e_q$ is the electric charge of the quark. $A_{\mu}$ is the vector potential of the SM photon. $F^{\prime}_{\mu\nu}$, $F_{\mu\nu}$ are the field strength tensor of the hidden and SM photons, respectively. $m_{\chi}$ is the MCP's mass. $\epsilon$ is the kinetic mixing parameter between the SM and hidden photons. The covariant derivative is
\begin{center}
\begin{equation}
D_{\mu}=\partial_{\mu}-ig_{\chi}A^{\prime}_{\mu}
\end{equation}
\end{center}
where $g_{\chi}$ is the gauge coupling of the U(1)$^{\prime}$ and $A^{\prime}_{\mu}$ is the vector potential of the hidden photon.
\par
Then we may calculate the cross sections for scattering of MCP on an isoscalar nucleon target N=(p+n)/2 at high energies. Those DIS cross sections mainly depends on the behavior of structure functions at small x, which is the Bjorken scaling parameter. Since the MCP-mediator coupling is equal to $\epsilon^2\alpha$, the DIS cross section of MCPs on nuclei is equivalent to $\epsilon^2$ times as much as that of electrons on nuclei via a NC interaction under electromagnetism, that is
\begin{center}
\begin{equation}
\sigma_{\chi N}\approx\epsilon^2\sigma^{\gamma}_{eN}
\end{equation}
\end{center}
where $\chi$ denotes a MCP with $\epsilon e$, N is a nucleon. $\sigma^{\gamma}_{eN}$ is the cross section depending on $\gamma$ exchange between elections and nuclei. That electron-nuclei cross section can be obtained by integrating over the following doubly differential cross section may be expressed in term of the structure functions as
\begin{center}
\begin{equation}
\frac{d^2\sigma^{\gamma}_{eN}}{dxdQ^2}=\frac{2\pi\alpha^2}{xQ^4}[Y_+\tilde{F}_2(x,Q^2)-y^2\tilde{F}_L(x,Q^2)]
\end{equation}
\end{center}
where $Q^2$ is the momentum transfer, $\alpha$ is the fine-structure constant. $Y_+=1+(1-y)^2$, the inelasticity parameter $y=\displaystyle\frac{Q^2}{2m_NE_{in}}$. $\tilde{F}_2(x,Q^2)$ and $\tilde{F}_L(x,Q^2)$ are the generalized structure functions which depend on $\gamma$ exchange between the electrons and nuclei. $m_N$ is the nucleon mass, $E_{in}$ is the incident electron energy (also the incident MCP energy). According to Next-to-leading (NLO) order QCD calculations, since, the contribution of the longitudinal structure function $\tilde{F}_L(x,Q^2)$ to that cross section is less than 1\%\cite{KR}, $\tilde{F}_L(x,Q^2)$ is ignored in this work. The $\tilde{F}_2(x,Q^2)$ term under electromagnetism is equal to a term depending on $\gamma$ exchange ($F_2^{\gamma}$), that is
\begin{center}
\begin{equation}
\tilde{F}_2=F_2^{\gamma}
\end{equation}
\end{center}
The structure function $F_2^{\gamma}$ can be expressed in terms of the quark and anti-quark parton distribution functions (PDFs) as
\begin{center}
\begin{equation}
F_2^{\gamma}=\sum_qe_q^2x(q+\bar{q})
\end{equation}
\end{center}
where the sum runs over quark flavors except the top quark (it is too massive to contribute significantly in the region of interest). A set of PDFs was determined with the LHC run II data\cite{ABMP}. This set of PDFs was taken to calculate the cross section for scattering of MCPs on nuclei in this work. For the PDFs of sea quarks, here, $s_s=\bar{s}_s=c_s=\bar{c}_s=b_s=\bar{b}_s$ was assumed in the calculation of those cross sections.
\par
The total DIS cross sections of MCPs on nuclei may be obtained through integrating over Eqn. (4) and calculating Eqn. (3). Their results can be approximately expressed as a simple power-law form in the energy range 1 TeV-10 PeV
\begin{center}
\begin{equation}
\sigma_{\chi N}\approx1.756\times10^{-31}\epsilon^2 cm^2 \left(\frac{E_{\chi}}{1GeV}\right)^{0.179}
\end{equation}
\end{center}
where E$_{\chi}$ is the MCP energy.
\par
The DIS cross-section for neutrino interaction with nuclei is computed in the lab-frame and given by simple power-law forms\cite{BHM} for neutrino energies above 1 TeV:
\begin{center}
\begin{equation}
\sigma_{\nu N}(CC)=4.74\times10^{-35} cm^2 \left(\frac{E_{\nu}}{1 GeV}\right)^{0.251}
\end{equation}
\end{center}
\par
\begin{center}
\begin{equation}
\sigma_{\nu N}(NC)=1.80\times10^{-35} cm^2 \left(\frac{E_{\nu}}{1 GeV}\right)^{0.256}
\end{equation}
\end{center}
where $\sigma_{\nu N}(CC)$ and $\sigma_{\nu N}(NC)$ are the DIS cross sections for neutrino scattering on nuclei via the charge current (CC) and neutral current (NC) interactions, respectively. $E_{\nu}$ is the neutrino energy.
\par
The inelasticity parameter $y=1 - \displaystyle\frac{E_{\chi^{\prime},lepton}}{E_{in}}$ (where $E_{in}$ is the incident MCP or neutrino energy and $E_{\chi^{\prime},lepton}$ is the outgoing MCPs or lepton energy). $E_{sec}=yE_{in}$, where $E_{sec}$ is the secondaries' energy after a MCP or neutrino interaction with nuclei. The mean values of $y$ for MCPs have been computed:
\begin{center}
\begin{equation}
\left\langle y \right\rangle =\frac{1}{\sigma(E_{in})} \int^1_0 y \frac{d\sigma}{dy}(E_{in},y)dy
\end{equation}
\end{center}
The MCP and neutrino interaction lengths can be obtained by
\begin{center}
\begin{equation}
L_{\nu,\chi}=\frac{1}{N_A\rho\sigma_{\nu,\chi N}}
\end{equation}
\end{center}
where $N_A$ is the Avogadro constant, and $\rho$ is the density of matter, which MCPs and neutrinos interact with.
\section{Evaluation of the numbers of expected MCPs and neutrinos at IceCube}
The IceCube detector is deployed in the deep ice below the geographic South Pole\cite{icecube2004}. It can detect neutrino interactions with nuclei via the measurement of the cascades caused by their secondary particles above the energy threshold of 100 GeV\cite{icecube2014a}. The MCPs which pass through the IceCube detector would interact with the nuclei inside IceCube. This is similar to the NC DIS of neutrino interaction with nuclei, whose secondary particles would develop into a cascade at IceCube.
\par
MCP events were selected with the following event selection criteria in this analysis. First, only cascade events were kept. To reduce more background events initiated by atmospheric muon, Second, only up-going events occurring during a period in which the Sun was below the horizon were kept. Besides, only those up-going events from the Sun's direction were kept.
\par
The $C_1$ and $C_2$ factors should be considered in the evaluation of the numbers of expected MCPs. $C_1$ is equal to 68.3\% (that is 68.3\% of the MCP events reconstructed with IceCube fall into a window caused by one standard energy uncertainty). $C_2$ is equal to 50\% (that is 50\% of the MCP events reconstructed with IceCube fall into a window caused by one median angular uncertainty). Then the number N$_{det}$ of expected MCPs obeys the following equation:
\begin{center}
\begin{equation}
\frac{dN_{det}}{dE} =C_1\times C_2\times\int_T A_{eff}(E)\Phi_{MCP} P(E,\xi(t)) dt
\end{equation}
\end{center}
where $A_{eff}(E)$ obtained from the Fig. 2 in Ref.\cite{icecube2014a} is denoting the effective observational area for IceCube. E is denoting the energy of an incident particle. $P(E,\xi(t))$ can be given by the following equation:
\begin{center}
\begin{equation}
P(E,\xi(t))=exp(-\displaystyle\frac{D_e(\xi(t))}{L_{earth}})\left(1-exp(-\displaystyle\frac{D}{L_{ice}})\right).
\end{equation}
\end{center}
where $L_{earth,ice}$ is denoting the MCP interaction lengths with the Earth and ice, respectively. D is denoting the effective length in the IceCube detector and taken to be 1 km in this work. $D_e(\xi(t))=2R_esin(\xi(t))$ is denoting the distance through the Earth. R$_e$ is denoting the radius of the Earth. $\xi(t)$ is denoting the obliquity of the ecliptic changing with time. The maximum value of $\xi$ is 23.44$^{\circ}$.
\par
After rejecting track-like events, the background remains two sources: astrophysical and atmospheric neutrinos which pass through the detector of IceCube. Only a neural current interaction with nuclei is relevant to muon neutrinos considered here. The astrophysical neutrinos flux can be described by\cite{icecube2021a}
\begin{center}
\begin{equation}
\Phi_{\nu}^{astro}=\Phi_{astro}\times\left(\displaystyle\frac{E_{\nu}}{100TeV}\right)^{-(\alpha+\beta log_{10}(\frac{E_{\nu}}{100TeV}))}\times10^{-18}GeV^{-1} cm^{-2}s^{-1}sr^{-1}
\end{equation}
\end{center}
where $\Phi_{\nu}^{astro}$ is denoting the total astrophysical neutrino flux. The coefficients, $\Phi_{astro}$, $\alpha$ and $\beta$ are given in Fig. VI.10 in Ref.\cite{icecube2021a}. The atmospheric neutrinos flux can be described by\cite{SMS}
\begin{center}
\begin{equation}
\Phi_{\nu}^{atm} = C_{\nu}\left(\displaystyle\frac{E_{\nu}}{1GeV}\right)^{-(\gamma_0+\gamma_1x+\gamma_2x^2)}GeV^{-1} cm^{-2}s^{-1}sr^{-1}
\end{equation}
\end{center}
where $x=log_{10}(E_{\nu}/1GeV)$. $\Phi_{\nu}^{atm}$ is denoting the atmospheric neutrino flux. The coefficients, $C_{\nu}$ ($\gamma_0$, $\gamma_1$ and $\gamma_2$) are given in Table III in Ref.\cite{SMS}.
\par
The neutrinos fallen into the energy and angular windows mentioned above would also be regarded as signal candidate events, so the evaluation of the number of expected neutrinos has to be performed by integrating over the region caused by these windows. Then the number of expected neutrinos N$_{\nu}$ obeys the following equation:
\begin{center}
\begin{equation}
\frac{dN_{\nu}}{dE} =\int_T \int_{\theta_{min}}^{\theta_{max}} A_{eff}(E)(\Phi_{\nu}^{astro}+\Phi_{\nu}^{atm}) P(E,\xi(t),\theta)\frac{2\pi r_e(\xi(t))^2 sin2\theta}{D_e'(\epsilon(t),\theta)^2} d\theta dt
\end{equation}
\end{center}
where $r_e(\xi(t))=\displaystyle\frac{D_e(\xi(t))}{2}$. $\theta$ is denoting the angular separation between the neutrinos and the Sun's diretion. $\theta_{min}$ = 0 and $\theta_{max}$ = $\sigma_{\theta}$. $\sigma_{\theta}$ is denoting the median angular uncertainty for cascades at IceCube. The standard energy and median angular uncertainties can be obtained from the Ref.\cite{icecube2021ICRC} and Ref.\cite{icecube2013}, respectively. $P(E,\xi(t),\theta)$ can be given by
\begin{center}
\begin{equation}
P(E,\xi(t),\theta)=exp(-\displaystyle\frac{D_e'(\xi(t),\theta)}{L_{earth}})\left(1-exp(-\displaystyle\frac{D}{L_{ice}})\right)
\end{equation}
\end{center}
where $D_e'(\xi(t),\theta)=D_e(\xi(t))cos(\theta)$ is denoting the distance through the Earth.
\section{Results}
The distributions and numbers of expected MCPs and neutrinos were evaluated in the secondaries' energy range 100 GeV-100 TeV assuming 6 years of IceCube data. Fig. 1 shows the distributions with an energy bin of 100 GeV of expected MCPs and neutrinos. Compared to MCPs with $\epsilon^2$=10$^{-10}$ and $\tau_{\phi} = 10^{19}$ s, the numbers of neutrino events per energy bin are at least smaller by 4 orders of magnitude in the energy range 100 GeV-100 TeV. As shown in Fig. 1, the dominant background is caused by atmospheric neutrinos at energies below 5 TeV but astrophysical neutrinos at energies above about 10 TeV in this measurement.
\par
The numbers of expected neutrinos (see black dash line) are shown in Fig. 2. The evaluation of the numbers of expected neutrinos was performed through integrating over the region caused by the energy and angular windows described above. The black dot line denotes the number of expected atmospheric neutrinos. This figure indicates the neutrino background can be ignored at the secondary energies above 300 GeV in this measurement. The numbers of expected MCPs with $\epsilon^2=10^{-8}$ and $\tau_{\phi} = 10^{19}$ s can reach about 386 and 1 at 100 GeV and 65 TeV at IceCube, respectively, as shown in Fig. 2 (see the red solid line). Fig. 2 also presents MCPs with $\epsilon^2=10^{-9}$ (see the blue dash line) and $\epsilon^2=10^{-10}$(see the green dot line) could be detected below about 5 TeV and 380 GeV at IceCube, respectively, when $\tau_{\phi} = 10^{19}$ s.
\section{Discussion and Conclusion}
Ref.\cite{icecube2021b} presents an analysis of neutrino signals due to the DM annihilation in the Sun with 6 years of IceCube data. This analysis has not found any significant indication of neutrinos due to the DM annihilation in the Sun. Since the MCP and neutrino signals are hard to distinguish at IceCube, it is a reasonable assumption that no events are observed in the measurement of MCPs due to the decay of $\phi$ in the Sun at IceCube in 6 years. The corresponding upper limit on MCP flux at 90\% C.L. was calculated with the Feldman-Cousins approach\cite{FC} (see the black solid line in Fig. 3). Fig. 3 also presents the fluxes of expected MCPs with $\epsilon^2=10^{-8}$ (red solid line), 10$^{-9}$ (blue dash line) and 10$^{-10}$ (green dot line). That limit excludes the MCP fluxes with $\epsilon^2=10^{-8}$, $10^{-9}$ and $10^{-10}$  below about 25 TeV, 1.8 TeV and 200 GeV, respectively.
\par
With $\epsilon^2$ = 10$^{-8}$, 10$^{-9}$ and 10$^{-10}$, hence, the MCPs from the Sun can be measured in the energy ranges 25-65 TeV, 1.8-5 TeV and 200-380 GeV at IceCube, respectively, when $\tau_{\phi} = 10^{19}$ s. Based on the results described above, it is a reasonable conclusion that those MCPs could be directly detected in the energy range O(100GeV)-O(10TeV) at IceCube when $\epsilon^2\gtrsim10^{-10}$. Since these constraints are only given by the assumptions mentioned above, certainly, the experimental collaborations, like the IceCube collaboration, should be encouraged to conduct an unbiased analysis with the data of IceCube.
\par
Since $\Phi_{MCP}$ is proportional to $\displaystyle\frac{1}{\tau_{\phi}}$ (see Eqn. (3)), the above results actually depends on the lifetime of heavy DM, $\tau_{\phi}$. If $\tau_{\phi}$ varies from 10$^{20}$ s to 10$^{21}$ s, the numbers of expected MCPs with IceCube are less by from 1 to 2 orders of magnitude than that with $\tau_{\phi}=10^{19}$ s, respectively.
\par
Likewise, the upper limit for $\epsilon^2$ at 90\% C.L. can be calculated with the Feldman-Cousins approach. Fig. 4 shows these limits with $\tau_{\phi}$ = 10$^{19}$ s (see red solid line), 10$^{20}$ s (see blue dash line) and 10$^{21}$ s (see green dot line), respectively. If the heavy DM mass, $m_{\phi}$, is equal to 3 TeV (the corresponding MCP energy is just 1.5 TeV), as shown in Fig. 4, the region of $\epsilon^2>3.5\times10^{-11}$ (that is $\epsilon$ > $5.9\times10^{-6}$) is ruled out when $\tau_{\phi}=10^{19}$ s.
\par
The MCP mass, $m_{MCP}$, is at least taken to be less than 10 MeV, since it is assumed that the MCP mass are much less than that of a proton, as mentioned in Sec. 1. So the region of $\epsilon > 5.9\times10^{-6}$ is ruled out at 90\% C.L. in the $m_{MCP}$-$\epsilon$ plane, when $m_{MCP} < $ 10 MeV. This result is shown in Fig. 5. To compare to other observations on MCPs, this figure also shows the $\epsilon$ bounds from cosmological and astrophysical observations\cite{CM,DHR,DGR,JR}, accelerator and fixed-target experiments\cite{DCB,SLAC}, experiments for decay of ortho-positronium\cite{OP} and Lamb shift\cite{Lamb}. A new region of 0.6 MeV < $m_{MCP}$ < 10 MeV and $6\times10^{-6}$ < $\epsilon$ $\lesssim$ $10^{-4}$ is ruled out in the $m_{MCP}$-$\epsilon$ plane with 6 year of IceCube data, as shown in Fig. 5.
\par
The MCPs from the Sun's core could be more easily detected with IceCube, compared to those from the Earth's core (although it is closer to the IceCube detector than the Sun), since the $\phi$ accumulation in the Sun is much greater than that in the Earth\cite{BCH}. The numbers of expected MCPs in the case of the Earth are less by about 2 orders of magnitude than those in the case of the Sun, as I roughly evaluated them. The region of $\epsilon \gtrsim 10^{-3}$ is ruled out in the case of the Earth when $\tau_{\phi}=10^{19}$ s. Meanwhile, the numbers of expected MCPs from the decay of the galactic and extra-galactic $\phi$'s were roughly evaluated at IceCube. They are less by about 2 times than that in the case of the Sun. The $\epsilon$ below limit in the galactic and extra-galactic case is about 10$^{-5}$ when $\tau_{\phi}=10^{19}$ s.
\par
Since the decay of $\phi$'s into MCPs can lead to extra energy injection during recombination and reionization eras in the early universe, the parameters in this DM scenario may be constrained by early universe observations. Since the $\phi$ lifetime is much greater than the age of the Universe, however, $\Omega_{MCPs} h^2 \lesssim 10^{-12}\Omega_{DM}h^2$ in this scenario. Ref.\cite{DDRT} presented that the cosmological abundance of MCPs was strongly constrained by the Planck data, that was $\Omega_{MCPs}h^2<0.001$. I also arrived at the upper limit of $\epsilon \gtrsim 10^{-6}$ with the Planck data when $m_{MCP}$ = 1 MeV, according to Ref.\cite{DDRT}. This is in consistence with my result mentioned above. Thus, the parameters in this DM scenario can't be constrained by the present early universe observations.
\section{Acknowledgements}
This work was supported by the National Natural Science Foundation
of China (NSFC) under the contract No. 11235006, the Science Fund of
Fujian University of Technology under the contracts No. GY-Z14061 and GY-Z13114 and the Natural Science Foundation of
Fujian Province in China under the contract No. 2015J01577.
\par

\newpage

\begin{figure}
 \centering
 \includegraphics[width=0.9\textwidth]{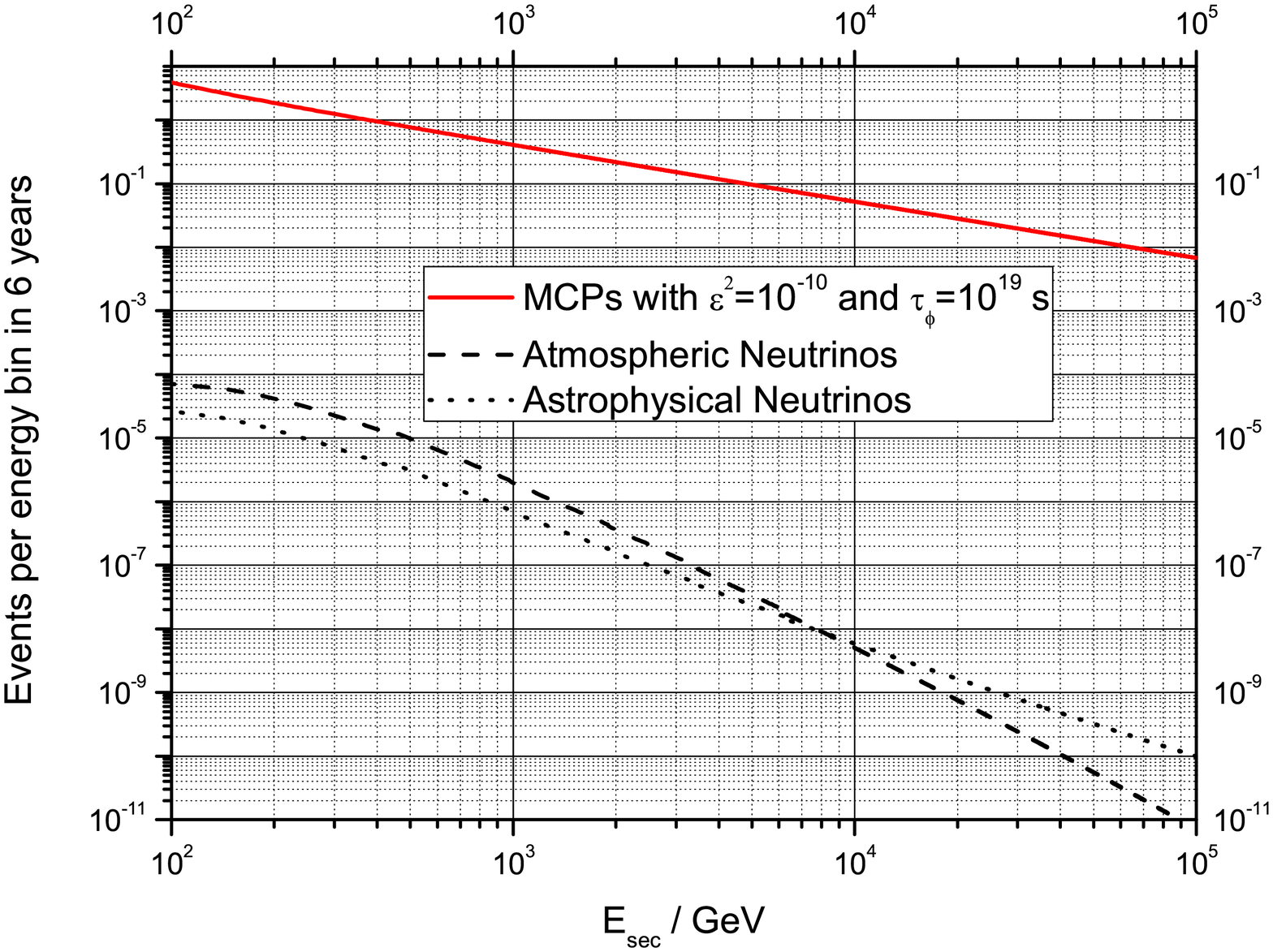}
 \caption{Distributions of expected MCPs with $\tau_{\phi}$ = $10^{19}$ s and $\epsilon^2=10^{-10}$ and astrophysical and atmospheric neutrinos. Their energy bins are 100 GeV.}
 \label{fig:E_bin}
\end{figure}

\begin{figure}
 \centering
 \includegraphics[width=0.9\textwidth]{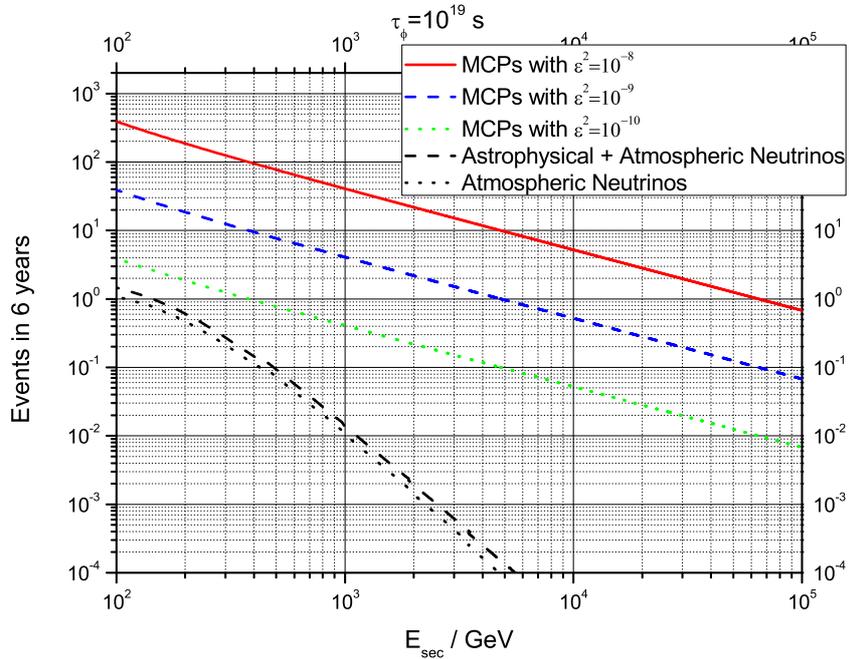}
 \caption{With the different $\epsilon^2$ (= $10^{-10}$, $10^{-9}$ and $10^{-8}$), the numbers of expected MCPs were evaluated assuming 6 years of IceCube data, respectively. The evaluation of numbers of expected neutrinos was performed by integrating over the regions caused by one standard energy and median angular uncertainties.}
 \label{fig:icecube_epsilon2_1e19}
\end{figure}

\begin{figure}
 \centering
 \includegraphics[width=0.9\textwidth]{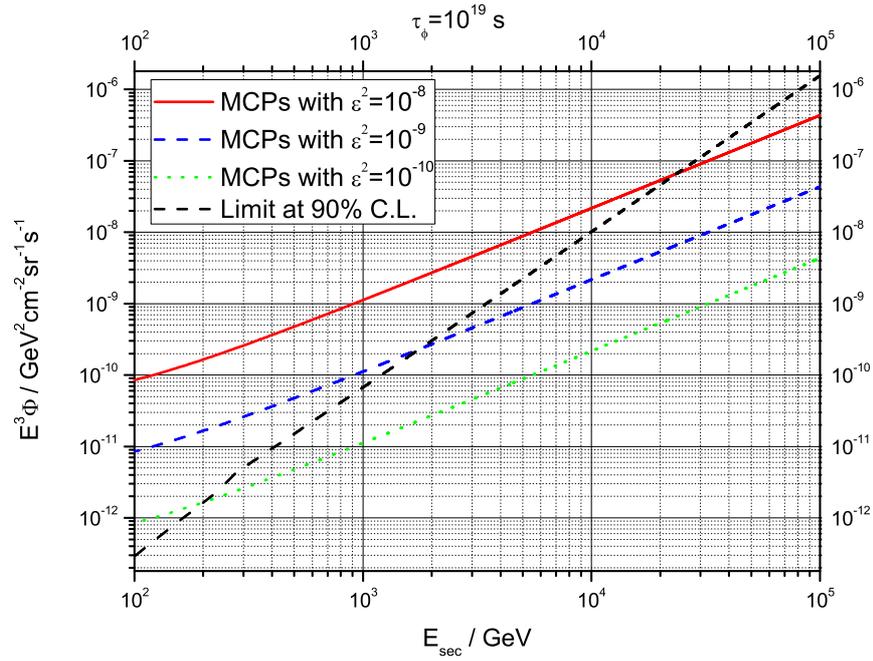}
 \caption{With the different $\epsilon^2$ (= $10^{-10}$, $10^{-9}$ and $10^{-8}$), the fluxes of expected MCPs were estimated at IceCube, respectively. Assuming no observation at IceCube in 6 years, the upper limit at 90\% C.L. was also computed.}
 \label{fig:flux}
\end{figure}

\begin{figure}
 \centering
 \includegraphics[width=0.9\textwidth]{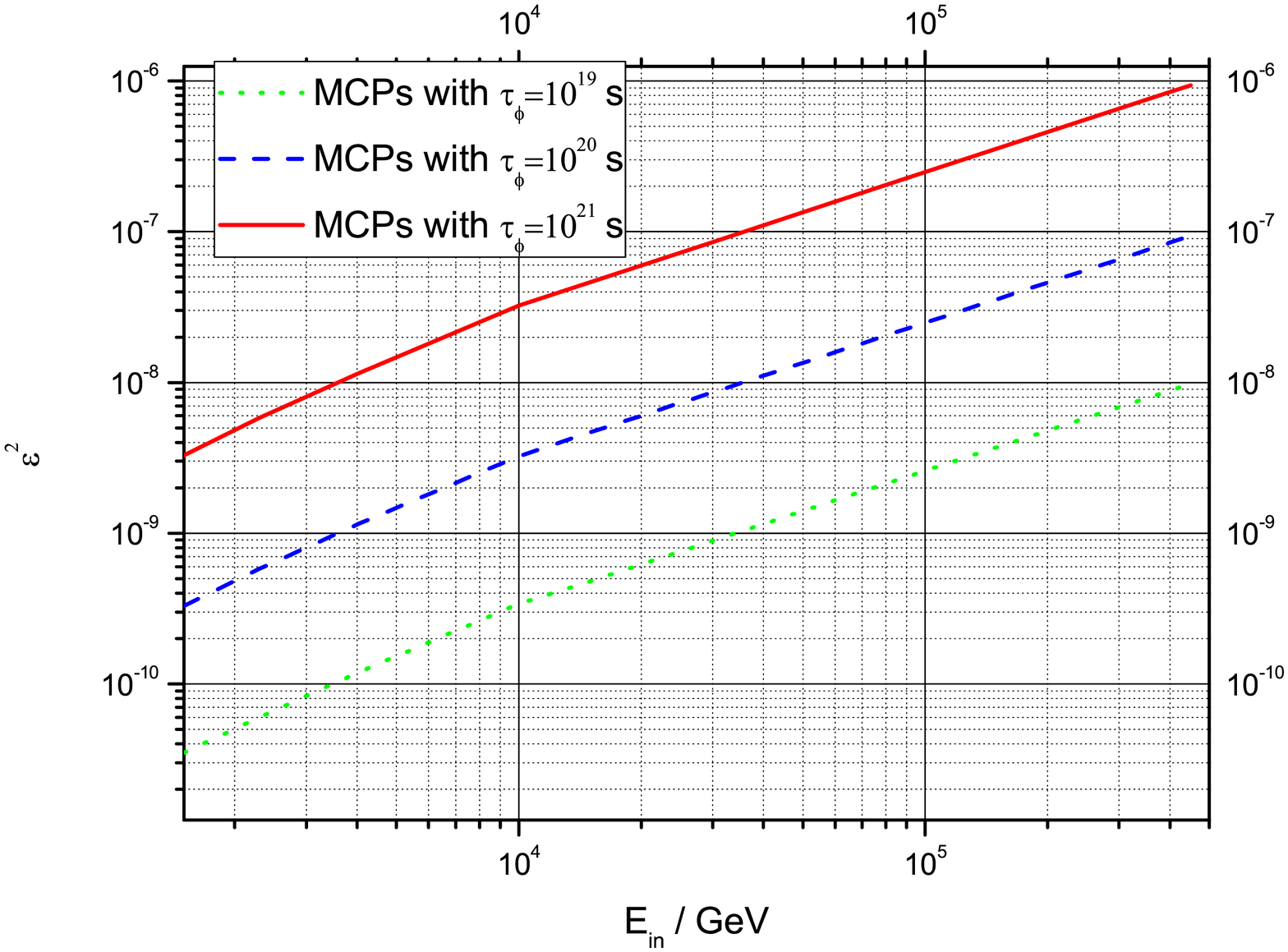}
 \caption{With the different $\tau_{\phi}$ (= $10^{19}$ s, $10^{20}$ s and $10^{21}$ s), the upper limit on $\epsilon$ at 90\% C.L. was computed, respectively, assuming no observation at IceCube in 6 years.}
 \label{fig:epsilon2}
\end{figure}

\begin{figure}
 \centering
 \includegraphics[width=0.9\textwidth]{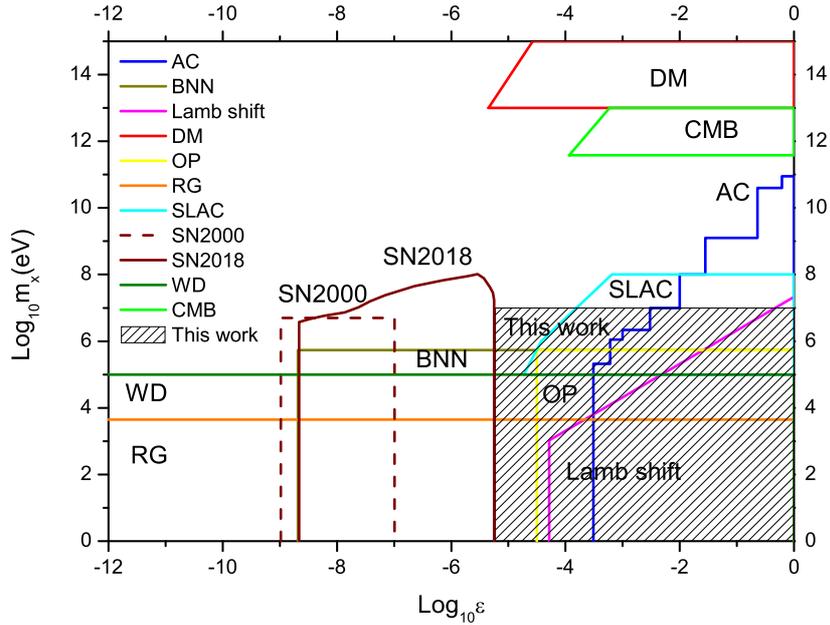}
 \caption{If $m_{\phi}$=3 TeV, with $\tau_{\phi}=10^{19}$ s, a new region (shaded region) is ruled out in the $m_{MCP}$ vs. $\epsilon$ plane, when $m_{MCP} < 10 MeV$ and $\epsilon > 5.9\times10^{-6}$ (this work). Meanwhile, the bounds from plasmon decay in red giants (RG)\cite{DHR}, plasmon decay in
white dwarfs (WD)\cite{DHR}, cooling of the Supernova 1987A (SN2000\cite{DHR}, SN2018\cite{CM}), accelerator (AC)\cite{DCB} and fixed-target experiments (SLAC)\cite{SLAC}, the Tokyo search for the invisible decay of ortho-positronium (OP)\cite{OP}, the Lamb shift\cite{Lamb}, big bang nucleosynthesis (BBN)\cite{DHR}, cosmic microwave background (CMB)\cite{DGR} and dark matter searches (DM)\cite{JR} are also plotted on this figure.}
 \label{fig:epsilon_bound}
\end{figure}

\end{document}